\documentclass[aps,prx,twocolumn,showpacs,superscriptaddress,floatfix]{revtex4-2}  
\usepackage{float}
\usepackage{bbm}
\usepackage{epsfig}
\usepackage{epstopdf}
\usepackage{graphicx}
\usepackage{amsmath,amssymb}
\usepackage{bm}
\usepackage{physics}
\usepackage{color}
\usepackage{lineno,blindtext}
\usepackage{siunitx, booktabs}
\usepackage{diagbox, eqparbox, hhline}
\usepackage{soul}
\usepackage{xcolor}

\usepackage[normalem]{ulem}
\newcolumntype{P}[1]{>{\centering\arraybackslash}p{#1}}
\usepackage[colorlinks,citecolor=blue,linkcolor=blue,urlcolor=blue,]{hyperref}
\usepackage{graphicx,dsfont,amssymb,booktabs}

\begin{document}

\title{Cooperative non-reciprocal  emission and quantum sensing of symmetry breaking}

\author{Xin Li}
\email{licqp@bc.edu}
\affiliation{Department of Physics, Boston College, 140 Commonwealth Avenue, Chestnut Hill, Massachusetts 02467, USA}

\author{Benedetta Flebus}
\affiliation{Department of Physics, Boston College, 140 Commonwealth Avenue, Chestnut Hill, Massachusetts 02467, USA}

\begin{abstract}

Non-reciprocal propagation of energy and information is fundamental to a wide range of quantum technology applications. In this work, we explore the quantum many-body dynamics of a qubit ensemble coupled to a shared bath that mediates coherent and dissipative inter-qubit interactions with both symmetric and anti-symmetric components. We find that the interplay between anti-symmetric (symmetric) coherent and symmetric (anti-symmetric) dissipative interactions results in non-reciprocal couplings, which, in turn, generate a spatially asymmetric emission pattern. We demonstrate that this pattern arises from non-reciprocal interactions coupling different quantum many-body states within a specific excitation manifold.
Focusing on solid-state baths, we show that their lack of time-reversal and inversion symmetry is a key ingredient for generating non-reciprocal dynamics in the qubit ensemble. With the plethora of quantum materials that exhibit this symmetry breaking at equilibrium, our approach paves the way for realizing cooperative non-reciprocal transport in qubit ensembles without requiring time-modulated external drives or complex engineering. Using an ensemble of nitrogen-vacancy (NV) centers coupled to a generic non-centrosymmetric ferromagnetic bath as a concrete example, we demonstrate that our predictions can be tested in near-future experiments. Additionally, we find that the spatial asymmetry in the relaxation dynamics increases with the size of the qubit ensemble and persists over longer inter-qubit distances than the signatures of super- and sub-radiant collective relaxation dynamics.
As the spatial asymmetry in the relaxation dynamics of the qubit ensemble is a direct probe of symmetry breaking in the solid-state bath, our work also opens the door to developing model-agnostic quantum sensing schemes capable of detecting  bath properties invisible to current state-of-the-art protocols, which operate solid-state defects as single-qubit sensors.

\end{abstract}

\maketitle

%%%%%%%%%%%%%%%%%%%%%%%%%%%%Main Body%%%%%%%%%%%%%%%%%%%%%%%%%%%%%%%%%%%%%

\section{Introduction}

Engineering non-reciprocal interactions among quantum systems is crucial for a wide range of emerging quantum technologies, from light harvesting~\cite{hertzog2019strong,wu2023nonreciprocal,zhang2022nonreciprocal} to quantum information processing~\cite{kimble2008quantum,hurst2018nonreciprocal,ren2022nonreciprocal,wang2023quantum}. Chiral light-matter interactions, for instance, have been a primary focus in the field of quantum optics due to their fundamental interest~\cite{wang2023quantum,metelmann2015nonreciprocal,clerk2022introduction,jen2020subradiance,ayuso2019synthetic,yoo2015chiral,lininger2023chirality,genet2022chiral} and their potential as building blocks for non-reciprocal devices and complex quantum networks \cite{stannigel2012driven,lodahl2017chiral,metelmann2018nonreciprocal,huang2018nonreciprocal}. These interactions naturally emerge in photonic nanostructures that strongly confine light, resulting in spin-momentum locking, where the coupling between light and quantum emitters depends on both the direction of light propagation and the polarization of the emitter. To date, several experiments using specially designed photonic crystal waveguides have demonstrated directional spontaneous photon emission driven by optical spin-momentum locking effect~\cite{sollner2015deterministic,mitsch2014quantum,lodahl2017chiral,kalhor2016universal,van2016universal}.

An alternative  route to non-reciprocal photon transmission was introduced by Ref.~\cite{metelmann2015nonreciprocal}, which identifies the breaking of time-reversal symmetry (TRS) in the reservoir, combined with a careful balancing of the bath-mediated coherent and dissipative inter-qubit interactions, as key ingredients for engineering quantum non-reciprocal dynamics. This proposal counts already experimental realizations~\cite{sliwa2015reconfigurable,fang2017generalized,lecocq2017nonreciprocal,ruesink2016nonreciprocity}, many of which employ time-modulated drives to break the effective TRS of the reservoir. Within this context,  a promising, and yet vastly unexplored, platform  is offered by quantum hybrid solid-state systems comprising an ensemble of solid-state spin defects interacting with the electromagnetic noise emitted by a common solid-state reservoir~\cite{li2023solid}. A clear advantage of this setup lies in the broad range of solid-state systems that display symmetry breaking at equilibrium and whose fluctuating spin, pseudospin, or charge degrees of freedom  can generate magnetic fields, which hints at the possibility of engineering quantum non-reciprocity without the need for time-dependent drives or complex microfabrication techniques~\cite{nagaosa2024nonreciprocal,cheong2018broken,tokura2018nonreciprocal}.
%As chirality has been shown to enhance the robustness of entanglement against experimental imperfections~\cite{gonzalez2015chiral}, this approach might pave the way for creating quantum networks based on solid-state spin defects, which remains an ongoing challenge.

 Recent work has shown that the coupling between an ensemble of solid-state spin defects and a shared magnetic reservoir can produce dissipative inter-qubit interactions strong enough to induce super- and sub-radiant dynamics~\cite{li2023solid}. Such interactions are not only potential sources of cooperative quantum behavior and long-range entanglement, but also encode information about the noise inaccessible to conventional quantum sensing schemes operating spin defects as single-qubit sensors~\cite{casola2018probing,dolde2013room,bermudez2011electron,gaebel2006room}. This naturally prompts the question of whether the relaxation dynamics of a qubit ensemble interacting dissipatively with a bath might exhibit qualitative signatures of the symmetries (or lack thereof) of the latter. If confirmed, this finding could drive the advancement of a novel, model-agnostic quantum sensing approach that leverages quantum correlations among multiple sensors.

Inspired by these considerations, here we explore the many-body quantum dynamics of a qubit ensemble whose interaction with a common reservoir enables both spatially symmetric and anti-symmetric coherent and dissipative inter-qubit couplings. In agreement with Ref.~\cite{metelmann2015nonreciprocal}, we find that the interplay between coherent anti-symmetric and dissipative symmetric interactions can generate non-reciprocal quantum dynamics. Interestingly, our analysis reveals that a balancing between coherent symmetric and dissipative anti-symmetric couplings can take place concurrently. Furthermore, we find that the spatial asymmetry of the relaxation dynamics of an ensemble of $N$ qubits can result from the coherent asymmetric interaction mixing states within a given excitation manifold and further prove our findings for a system of $N=2$ qubits.

\begin{figure}[t!]
\includegraphics[width=1\linewidth]{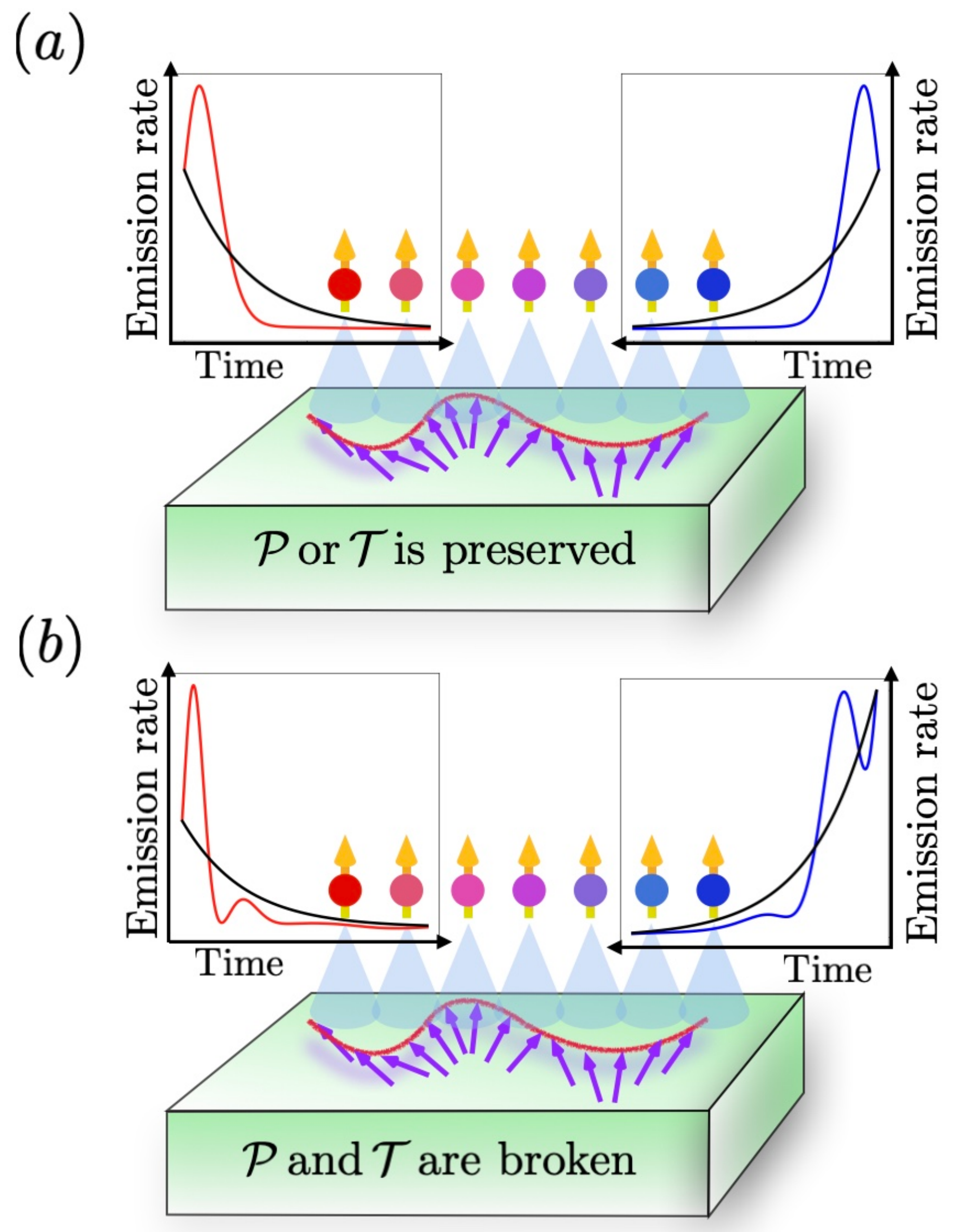}
    \caption{Schematic illustration of the emission pattern of a qubit array as a function of the bath symmetries. (a) The leftmost and rightmost qubits of an ensemble exhibit identical relaxation dynamics when coupled to a reservoir where either spatial inversion symmetry ($\mathcal{P}$) or time-reversal symmetry ($\mathcal{T}$) is preserved. (b) When both $\mathcal{P}$ and $\mathcal{T}$ are broken in the reservoir, the leftmost and rightmost qubits display distinct emission patterns. The red and blue curves represent the emission rates of the leftmost and rightmost qubits, respectively, while the black curve shows the exponentially decaying relaxation dynamics of an isolated qubit.}  
    \label{Fig1}
\end{figure}

Next, we specialize to solid-state reservoirs and  show  that their lack of effective time-reversal $\mathcal{T}$ and spatial inversion $\mathcal{P}$ symmetries  play a crucial role, i.e., a non-reciprocal bath response  is required for the generation of many-body quantum non-reciprocal dynamics.
As a concrete example, we consider an ensemble of NV centers interacting  with a  non-centrosymmetric magnetic reservoir and prove that it obeys the master equation that is at the center of our analysis. We calculate the emission rates for realistic experimental parameters and find that the spatial asymmetry of the sensor relaxation dynamics can serve as an indicator of symmetry breaking in the solid-state bath, as illustrated in Fig.~\ref{Fig1}. Notably, through our case study we also unveil the cooperative and long-range nature of the non-reciprocal  dissipative inter-qubit interactions mediated by the solid-state reservoir.

This work is organized as follows. In Sec.~\ref{model}, we set the stage for our investigation by introducing the master equation that is at the center of our work while uncovering its connection to the paradigmatic Hatano-Nelson model.  
In Sec.~\ref{nonreciprocalemi},  we show that the spatial asymmetry in the qubit relaxation dynamics arises from the interplay between coherent anti-symmetric (symmetric) and dissipative symmetric (anti-symmetric) inter-qubit interactions, and we further explore how these interactions shape the dynamical evolution of many-body states. In Sec.~\ref{SSbath}, we specialize to a qubit array interacting with the magnetic field fluctuations emitted by a common solid-state reservoir and prove that both time-reversal and inversion symmetries must be broken within the reservoir for non-reciprocal inter-qubit interactions to emerge. In Sec.~\ref{concretefeeo}, we apply our framework to a realistic experimental setup, i.e., an ensemble of NV centers interacting with a common non-centrosymmetric ferromagnetic bath, and investigate the dependence of the qubit relaxation dynamics on experimentally tunable parameters. Finally, in Sec.~\ref{conclusion}, we provide a summary and outlook.

\section{MODEL \label{model}}

In this Section, we introduce the model that is at the center of our investigation, i.e., a one-dimensional (1$d$) array of $N$ two-level qubits coupled to a common reservoir that mediates both symmetric and anti-symmetric inter-qubit interactions. Specifically, we assume that, upon tracing out the reservoir degrees of freedom under a Born-Markov approximation~\cite{breuer2002theory,manzano2020short}, the time evolution of the density matrix of the qubit ensemble is given by
\begin{align}
    \frac{d\rho}{dt}
    &=-i[\mathcal{H},\rho]+\mathcal{L}[\rho]\,,\label{masterfull}
\end{align}
where the Hamiltonian $\mathcal{H}$ reads as
\begin{align}
\mathcal{H}=\frac{1}{2}\sum_{\alpha,\beta=1}^NJ_{\alpha\beta}\sigma_\alpha^+\sigma_\beta^-\,,
\label{Hmasterfull}
\end{align}
and 
\begin{align}
\mathcal{L}[\rho] =\sum_{\alpha,\beta=1}^N\gamma_{\alpha\beta}\left(\sigma_\beta^-\rho\sigma_\alpha^+-\frac{1}{2}\{\sigma_\alpha^+\sigma_\beta^-,\rho\}\right)\,
\label{Lmasterfull}
\end{align}
is the Lindbladian superoperator. Here  $\sigma_\alpha^+=\ket{e_\alpha}\bra{g_\alpha}$ and $\sigma_\alpha^-=\ket{g_\alpha}\bra{e_\alpha}$ are, respectively, the raising and lowering operators describing the transition between the excited, i.e., $\ket{e_\alpha}$,  and the ground, i.e., $\ket{g_\alpha}$, states of the qubit $\boldsymbol{\sigma}_{\alpha}$ residing at the site $\mathbf{r}_{\alpha}$.  The coefficients $J_{\alpha\beta}$ and $\gamma_{\alpha\beta}$ parameterize, respectively, the coherent and dissipative interactions between the $\alpha$th and $\beta$th qubits. 

In general, both $J_{\alpha\beta}$ and $\gamma_{\alpha\beta}$ can be complex, while the hermiticity relations $J_{\alpha\beta}=J^*_{\beta\alpha}$ and $\gamma_{\alpha\beta}=\gamma^*_{\beta\alpha}$ imply that their real and imaginary parts are, respectively, symmetric and anti-symmetric upon the exchange of the indices $\alpha$ and $\beta$. Thus, one can write $J_{\alpha\beta}=J^s_{\alpha\beta}+iJ^a_{\alpha\beta}$ and $\gamma_{\alpha\beta}=\gamma^s_{\alpha\beta}+i\gamma^a_{\alpha\beta}$, with $J^s_{\alpha\beta}=J^s_{\beta\alpha}$, $\gamma^s_{\alpha\beta}=\gamma^s_{\beta\alpha}$, and $J^a_{\alpha\beta}=-J^a_{\beta\alpha}$, $\gamma^a_{\alpha\beta}=-\gamma^a_{\beta\alpha}$. 
To visualize the potential source of quantum non-reciprocity in our model, it is helpful to rewrite Eqs.~\eqref{masterfull}-\eqref{Lmasterfull} as
\begin{align}
    \frac{d\rho}{dt}
    &=-i\left(\mathcal{H}_{eff}\rho-\rho\mathcal{H}^\dag_{eff}\right)+\sum_{\alpha,\beta=1}^N\gamma_{\alpha\beta}\sigma_\beta^-\rho\sigma_\alpha^+\,,\label{EffLindbladian}
\end{align}
where 
\begin{align}
 \mathcal{H}_{eff}
=\sum^N_{\substack{\alpha,\beta=1 \\  \alpha<\beta}}(\gamma^{L}_{\alpha\beta}\sigma_\alpha^+\sigma_\beta^-+\gamma^{R}_{\alpha\beta} \sigma_\beta^+\sigma_\alpha^-)\,\label{complexnonher}
 \end{align}
takes a form similar to the Hatano-Nelson model~\cite{hatano1996localization,hatano1997vortex,okuma2023non}, a paradigmatic example of a non-Hermitian Hamiltonian displaying the non-Hermitian skin effect~\cite{deng2022non,hurst2022non,zhang2022review,okuma2023non}.
 Here,  we have defined the local dissipation as $ \gamma_0 \equiv \gamma_{\alpha \alpha}$ and neglected the on-site interactions $J_{\alpha\alpha}$ since they do not play a significant role in our analysis. The complex right-to-left and left-to-right hoppings are given, respectively, by
\begin{align}
\gamma^{L}_{\alpha\beta}&=\frac{1}{2}\left[\left(J^s_{\alpha\beta}+\gamma^a_{\alpha\beta}\right)+i\left(J^a_{\alpha\beta}-\gamma^s_{\alpha\beta}\right)\right]\,,\label{leftcoe}\\
\gamma^{R}_{\alpha\beta}&=\frac{1}{2}\left[\left(J^s_{\alpha\beta}-\gamma^a_{\alpha\beta}\right)-i\left(J^a_{\alpha\beta}+\gamma^s_{\alpha\beta}\right)\right]\,.\label{rightcoe}
 \end{align}
Equations (\ref{leftcoe}) and (\ref{rightcoe}) show that, in the absence of anti-symmetric interactions, i.e., for $J^a_{\alpha\beta},\gamma^a_{\alpha\beta} = 0$, the inter-qubit interaction is reciprocal, i.e., $|\gamma^{R}_{\alpha\beta}| = |\gamma^{L}_{\alpha\beta}|$. The emergence of non-reciprocity in the ensemble dynamics, i.e., $|\gamma^{R}_{\alpha\beta}| \neq |\gamma^{L}_{\alpha\beta}|$, relies instead on the interplay between : i) the anti-symmetric  coherent, $J^{a}_{\alpha\beta}$, and symmetric  dissipative, $\gamma^{s}_{\alpha\beta}$, interactions, and ii)  the symmetric coherent, $J^{s}_{\alpha\beta}$, and anti-symmetric  dissipative, $\gamma^{a}_{\alpha\beta}$, interactions. When considering short-range (i.e., nearest-neighbors) interactions, the mechanism~(i) is formally analogous to the reservoir-engineering approach proposed in Ref.~\cite{metelmann2015nonreciprocal}, according to which a careful balancing between $J^{a}_{\alpha\beta}$ and $\gamma^{s}_{\alpha\beta}$, i.e., $J^a_{\alpha\beta}=\pm \gamma^s_{\alpha\beta}$, can enable unidirectional photon hopping. However, as we will discuss in detail in Sec.~\ref{SSbath} when considering a common solid-state reservoir, the Kramers-Kronig relations constrain the coherent and dissipative inter-qubit interactions to be interdependent. Consequently, as shown by Eqs.~(\ref{leftcoe}) and (\ref{rightcoe}), realizing the unidirectional inter-qubit couplings that characterize cascade quantum systems~\cite{gardiner1993driving,carmichael1993quantum,faist1994quantum,gmachl2001recent} requires a careful balancing of each hopping amplitude. Specifically, the unidirectionality condition  $\gamma_{\alpha\beta}^{L(R)}=0$ is satisfied only when both conditions $J^s_{\alpha\beta}=\mp\gamma^a_{\alpha\beta}$ and $ J^a_{\alpha\beta}=\pm\gamma^s_{\alpha\beta}$ are satisfied for any $\alpha \neq \beta$, which might be challenging to engineer, particularly for baths mediating long-range non-periodic inter-qubit interactions.

\section{Non-reciprocal emission \label{nonreciprocalemi}}

In this Section, we  explore the impact of the non-reciprocity encoded in Eqs.~\eqref{leftcoe} and~\eqref{rightcoe} on the qubit relaxation dynamics. We first develop insights for an ensemble with an arbitrary number  $N>1$ of qubits, which we then validate through the analytically solvable case of $N=2$ qubits.
To this end, we begin by writing  the emission rate $\mathcal{R}_{\alpha}(t)$ of the $\alpha$th qubit obeying the master equation~\eqref{EffLindbladian} as
 
 \begin{align}
     \mathcal{R}_\alpha(t)=-\frac{1}{2}\frac{d}{dt}\text{Tr}(\rho\sigma_\alpha^z)=\sum_{\substack{\beta=1 \\  \beta\neq\alpha}}^N \mathcal{R}_{\alpha \beta}(t)+\gamma_0\langle\sigma_\alpha^+\sigma_{\alpha}^-\rangle\,,\label{indiviemimod}
 \end{align} 
where the quantity

\begin{align}
\mathcal{R}_{\alpha\beta}(t)=
\begin{cases}
  i\gamma^L_{\alpha\beta}\langle\sigma_\alpha^+\sigma_{\beta}^-\rangle + h.c.\,,\quad \text{for}\,\beta>\alpha\,;\\      
  i\gamma^R_{\alpha\beta}\langle\sigma_\alpha^+\sigma_{\beta}^-\rangle+ h.c.\,,\quad \text{for}\,\beta<\alpha\,,\label{spatialemrate}
\end{cases}
\end{align}
can be understood as the contribution of the $\beta$th qubit dynamics to the emission rate of the $\alpha$th qubit. Here, $h.c.$ stands for the corresponding Hermitian conjugate components. Using Eq.~\eqref{spatialemrate}, we introduce a parameter that quantifies the degree of non-reciprocity in how a pair of qubits influences each other's dynamics, i.e.,
\begin{align}
\Delta\mathcal{R}_{\alpha\beta}(t)&=\mathcal{R}_{\alpha \beta }(t)-\mathcal{R}_{\beta\alpha }(t)\,, \nonumber \\
&=-2J^s_{\alpha\beta}\text{Im}\langle\sigma_\alpha^+\sigma_\beta^-\rangle-2J^a_{\alpha\beta}\text{Re}\langle\sigma_\alpha^+\sigma_\beta^-\rangle\,.
\label{259}
\end{align}
In the absence of anti-symmetric interactions, i.e., $J_{\alpha\beta}^a, \gamma_{\alpha\beta}^a=0$, the second term on the right-hand side of Eq.~\eqref{259} trivially vanishes.  The first term on the right-hand side of Eq.~\eqref{259} vanishes as well, since the two-point correlation only depends on the relative distance between two qubits, i.e., $\langle\sigma_\alpha^+\sigma^-_\beta\rangle = \langle\sigma_\beta^+\sigma^-_\alpha\rangle$ for $  N \rightarrow \infty$, which implies $\text{Im}\langle\sigma_\alpha^+\sigma_\beta^-\rangle=0$.  Conversely, when the strength of both symmetric (anti-symmetric) coherent, $J^{s(a)}_{\alpha\beta}$, and anti-symmetric (symmetric) dissipative, $\gamma^{a(s)}_{\alpha\beta}$, are finite,  Eq.~\eqref{259}  is non-vanishing, potentially resulting into non-reciprocal  relaxation dynamics.

\begin{figure}[t!]
\includegraphics[width=1\linewidth]{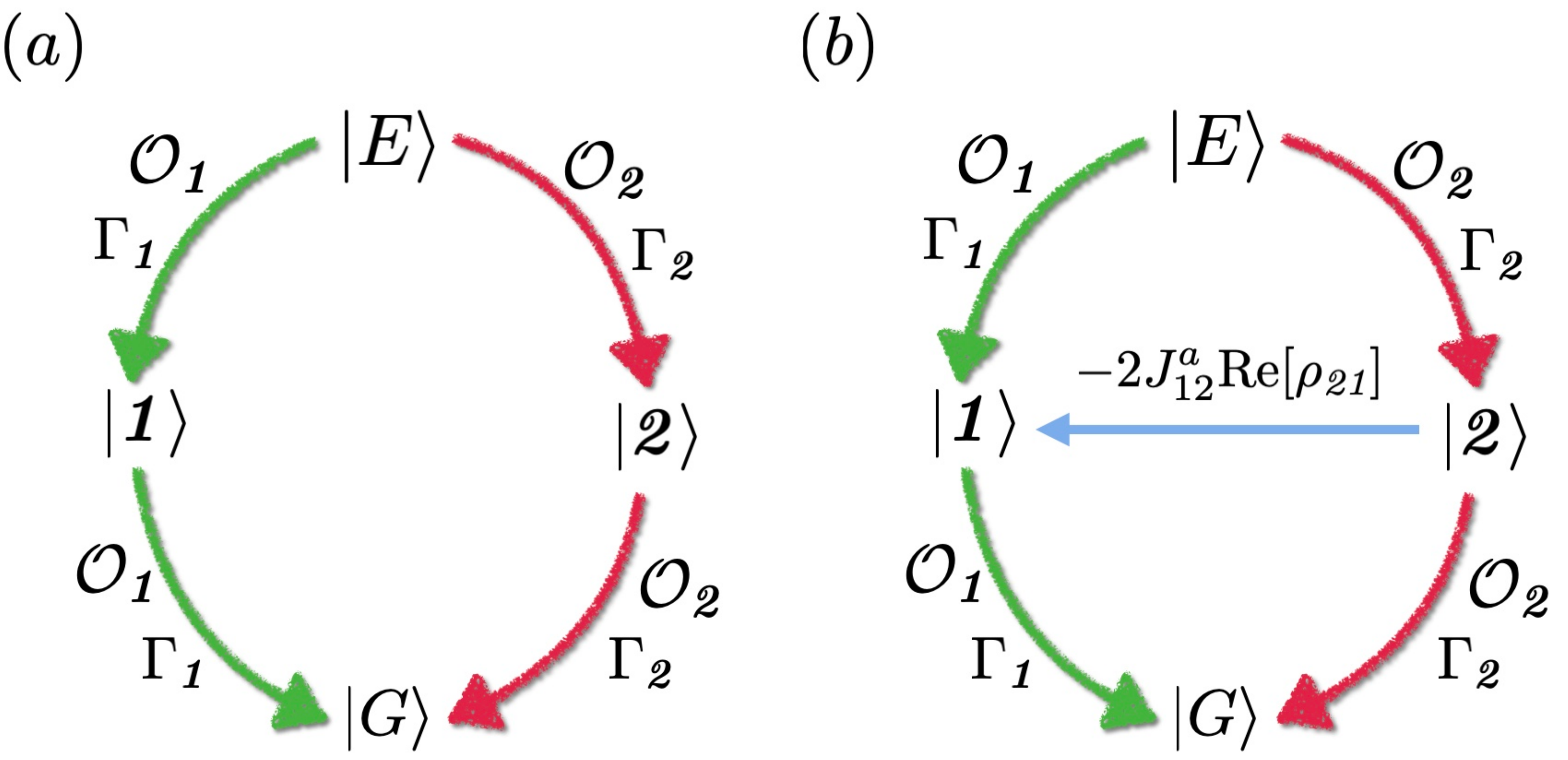}
    \caption{Schematic illustration of the role played by the anti-symmetric coherent interaction $J^a_{12}$ on the relaxation dynamics of two qubits. (a) In the absence of anti-symmetric interactions, the system decays from a symmetric (fully excited) state $\ket{E}$ to a symmetric (fully de-excited) state $\ket{G}$ along two independent trajectories. (b) The anti-symmetric interaction couples the singlet state $\ket{\textbf{\textit{1}}}$ and the triplet state $\ket{\textbf{\textit{2}}}$ in a non-reciprocal fashion - as a consequence, the previously independent decay paths become interdependent.}
    \label{Fig2}
\end{figure}

The non-reciprocal emission resulting from the interplay between coherent and dissipative interactions can be related also to the dynamical evolution of its many-body quantum states. To formalize this insight,
we rewrite the Lindbladian superoperator (\ref{Lmasterfull}) as
\begin{align}
    \mathcal{L}[\rho]&=\sum_{n=\textbf{\textit{1}}}^{N}\Gamma_{n}\left(\mathcal{O}_{n}\rho\mathcal{O}_{n}^\dag-\{\mathcal{O}_{n}^\dag\mathcal{O}_{n},\rho\}/2\right)\,,
    \label{Lrho}
\end{align}
where  $\mathcal{O}_{n}=\sum_{\alpha=1}^{N}S^\dag_{n\alpha}\sigma_\alpha^-$ is a collective jump operators and $\Gamma_{n}=\sum_{\alpha,\beta=1}^NS^\dag_{n\,\alpha}\gamma_{\alpha \beta}S_{\beta\,{n}}$ its decay rate, obtained, respectively, as an eigenvector and eigenvalue of the $N \times N$ decoherence matrix $\boldsymbol{\Gamma}= \left( \gamma_{\alpha \beta}\right)$. Here, $S$ is the unitary matrix that diagonalizes the decoherence matrix $\boldsymbol{\Gamma}$. In the basis of the collective jump operators, the Hermitian Hamiltonian~\eqref{Hmasterfull} reads as
\begin{align}
    \mathcal{H}=\frac{1}{2}\sum_{n,m=\textbf{\textit{1}}}^{N}J_{nm}\mathcal{O}_{n}^\dag\mathcal{O}_{m}\,,
    \label{Hjump}
\end{align}
where $J_{nm}=J^s_{nm}+iJ^a_{nm}$ consists of a Hermitian component, i.e., $(J^s_{nm})^*=J^s_{mn}$, and an anti-Hermitian component, $(J^a_{nm})^*=-J^a_{mn}$,  given as
\begin{align}
    J^s_{nm}&=\sum_{\alpha,\beta=1}^NS^\dag_{n\alpha}J^s_{\alpha\beta}S_{\beta\,m},\,\\
    J^a_{nm}&=\sum_{\alpha,\beta=1}^NS^\dag_{n\alpha}J^a_{\alpha\beta}S_{\beta\,m}\,.
\end{align}
Although solving the master equation analytically in the full Hilbert space is challenging, key insights can be gained by restricting the analysis to the eigenstates of the single-excitation manifold. Specifically, these eigenstates are given by $\ket{n} = \mathcal{O}^\dag_{n} \ket{G}$, where $n = \textbf{\textit{1}}, ..., N$, and $\ket{G}$ denotes the ground state.
We denote the probability for the qubit array to be in the state $\ket{n}$ as $\rho_{nn}=\bra{n}\rho\ket{n}$ and the coefficient of transitioning from the state $\ket{m}$ to $\ket{n}$ as $\rho_{n m}=\bra{n}\rho\ket{m}$.  The evolution of the probability $\rho_{nn}$ is governed by the following equation:
\begin{align}
    \frac{d \rho_{nn}}{dt}&=-i\bra{n}[\mathcal{H},\rho]\ket{n}+\bra{n}\mathcal{L}[\rho]\ket{n}\,,\label{probflow}
\end{align}
where the first and the second term on the right-hand side of Eq.~(\ref{probflow}) represent the contributions to the probability dynamics from, respectively, the Hermitian Hamiltonian $\mathcal{H}$ and the Lindbladian dissipator $\mathcal{L}[\rho]$. Making use of the unitary property $\sum_{l=\textbf{\textit{1}}}^{N}S^\dag_{nl} S_{lm}=\delta_{nm}$ and the commutator $[\mathcal{O}^\dag_{n},\mathcal{O}_{m}]=\sum_{\alpha=1}^NS^{\dag}_{m\alpha}S_{\alpha\,n}\sigma^z_\alpha$, the first term of the right-hand side of Eq.~\eqref{probflow} can be written as 
\begin{align}
   -i\bra{n}[\mathcal{H},\rho]\ket{n}&=\sum_{m=\textbf{\textit{1}}}^{N}P_{nm}\,,\label{hamiltflow}
\end{align}
where
\begin{align}
    P_{nm}&=\text{Im}[J_{nm}\rho_{mn}]\,.\label{pairflow}
\end{align}
 The second term of the right-hand side of Eq.~\eqref{probflow} reads as   \begin{align}
    \bra{n}\mathcal{L}[\rho]\ket{n}&=\sum_{m=\textbf{\textit{1}}}^{N}\Gamma_{m}{\bra{nm}}\rho\ket{nm}-\Gamma_{n}\bra{n}\rho\ket{n},\label{lindbprobaflow}
\end{align}
where $\ket{nm}=\mathcal{O}^\dag_{n}\mathcal{O}^\dag_{m}\ket{G}$ denotes a state within the two-excitation manifold. 
Equations~(\ref{hamiltflow})–(\ref{lindbprobaflow}) show that the Hermitian Hamiltonian couples different collective states within the same excitation manifold, while the Lindbladian connects the state $\ket{n}$ to states in different manifolds via decay processes.
Next, we introduce a parameter that quantifies the degree of non-reciprocity of the probability flow between the two states $\ket{n}$ and $\ket{m}$ within the same excitation manifold as
\begin{align}
    \Delta P_{nm}&=P_{nm}-P_{mn},\nonumber\\
    &=2(\text{Re}[J_{nm}]\text{Im}[\rho_{mn}]+\text{Im}[J_{nm}]\text{Re}[\rho_{mn}])
    .\label{netproba}
\end{align}
In the absence of anti-symmetric interactions, i.e.,  $J^{a}_{\alpha\beta}, \gamma^{a}_{\alpha\beta}=0$, the unitary matrix $S$ used to diagonalize the decoherence matrix $\mathbf{\Gamma}$ is real and symmetric, i.e., $S^T=S$, where $T$ denotes the transpose operation. Under this condition, $J^s_{mn}$ is real, with $J^s_{mn}=\sum_{\alpha,\beta=1}^N S^T_{m\alpha} J^s_{\alpha\beta} S_{\beta,n}=(J^s_{mn})^*$, and $\rho_{nm}=\sum_{\alpha,\beta=1}^N S^T_{n\alpha} S_{\beta\,m} \bra{G}\sigma_\alpha^-\rho\sigma_\beta^+ \ket{G}$. In analogy with the analysis of Eq.~(\ref{259}), one finds $\bra{G}\sigma_\alpha^-\rho\sigma_\beta^+\ket{G}=\bra{G}\sigma_\beta^-\rho\sigma_\alpha^+\ket{G}$, which implies $\rho_{nm}$ is also real,  $\rho_{nm}=\rho_{mn}=(\rho_{nm})^*$. Thus, for a reciprocal system, we have $\Delta P_{nm}=0$, indicating that there is no non-reciprocal probability flow between states within the same excitation manifold. 
Otherwise, the interplay between  $J^a_{\alpha\beta}$ ($J^s_{\alpha\beta}$) and  $\gamma^s_{\alpha\beta}$ ($\gamma^a_{\alpha\beta}$) couples states within the same manifold, resulting in non-reciprocal emission. A comparison between Eqs.~\eqref{259} and~\eqref{netproba} suggests that the spatial asymmetry in qubit relaxation is linked to the non-reciprocal  dynamical evolution of the many-body quantum states. To solidify this intuition, we specialize to the case of
$N=2$ qubits and, for simplicity, focus on the non-reciprocity emerging from the interplay between the anti-symmetric coherent, i.e., $J^{a}_{\alpha\beta}$, and symmetric dissipative, i.e., $\gamma^{s}_{\alpha\beta}$, interactions. For a vanishing anti-symmetric component of the dissipative coupling, i.e., $\gamma^{a}_{\alpha\beta}=0$, the collective jump operators and the associated relaxation rates can be found as
\begin{align}
    \mathcal{O}_{\textbf{\textit{1}},\textbf{\textit{2}}}=\left(\sigma^-_2\mp\sigma^-_1\right)/\sqrt{2}\,,\quad \Gamma_{\textbf{\textit{1}},\textbf{\textit{2}}}=\gamma_0\mp\gamma^{s}_{12}\,.
\end{align}
Accordingly, the Hamiltonian~\eqref{Hjump} can be recast as
\begin{align}
\mathcal{H}&=\frac{1}{2}\sum_{l=\textbf{\textit{1}}} ^{\textbf{\textit{2}}}J^s_{ll}\mathcal{O}^\dag_{l}\mathcal{O}_{l}+\frac{i}{2} J_{\textbf{\textit{1}}\textbf{\textit{2}}}^a\left(\mathcal{O}^\dag_{\textbf{\textit{1}}}\mathcal{O}_{\textbf{\textit{2}}}-\mathcal{O}^\dag_{\textbf{\textit{2}}}\mathcal{O}_{\textbf{\textit{1}}}\right)\,,\label{Hamiltonew}
\end{align}
where  $J^s_{\textbf{\textit{1}}\textbf{\textit{1}(\textbf{\textit{2}}\textbf{\textit{2}})}}=\pm\,J^s_{12}$, and $J^a_{\textbf{\textit{1}}\textbf{\textit{2}}}=-J^a_{12}$.
Equation~\eqref{Hamiltonew} shows that,  unlike the symmetric coherent interaction $J^s_{12}$, the anti-symmetric coherent interaction $J^a_{12}$ can couple different collective jump operators within a given excitation manifold~\cite{zou2024spatially}. We will explicitly show that this coupling, when combined with dissipative interactions, is responsible for a non-reciprocal emission pattern by tracking the relaxation dynamics from the fully excited state $\ket{E}=\ket{e_1e_2}$. In the absence of anti-symmetric interactions, the excited state $\ket{E}$ decays sub-radiantly (super-radiantly) at a rate $\Gamma_{\textbf{\textit{1}}(\textbf{\textit{2}})}$ into the singlet (triplet) state $\ket{\textbf{\textit{1}},\textbf{\textit{2}}}=\mathcal{O}^\dag_{\textbf{\textit{1}},\textbf{\textit{2}}}\ket{G}=(|g_1e_2\rangle\mp|e_1g_2\rangle)/\sqrt{2}$, and then to the ground state $\ket{G}=\mathcal{O}_{\textbf{\textit{1}},\textbf{\textit{2}}}\ket{\textbf{\textit{1}},\textbf{\textit{2}}}=\ket{g_1g_2}$. Instead, when $J^a_{12}\neq 0$, these two independent decay channel can couple to each other, e.g., $iJ_{12}^a\mathcal{O}_{\textbf{\textit{1}}}^\dag\mathcal{O}_{\textbf{\textit{2}}}\ket{\textbf{\textit{2}}}=iJ_{12}^a\ket{\textbf{\textit{1}}}$,
as schematically illustrated in Fig.~\ref{Fig2}.
To clearly visualize this mechanism, we can derive the coupled equations describing the probability flow between the eigenstates from  Eq.~(\ref{probflow}) as

\begin{align}
    &\dot{\rho}_{EE}=-(\Gamma_{\textbf{\textit{1}}}+\Gamma_{\textbf{\textit{2}}})\rho_{EE}\,,\label{eqone}\\
    &\dot{\rho}_{\textbf{\textit{1}}\textbf{\textit{1}}}=-\Gamma_{\textbf{\textit{1}}}\rho_{\textbf{\textit{1}}\textbf{\textit{1}}}+\Gamma_{\textbf{\textit{1}}}\rho_{EE}-J_{12}^a\text{Re}[\rho_{\textbf{\textit{2}}\textbf{\textit{1}}}]\,,\label{singdmi}\\
    &\dot{\rho}_{\textbf{\textit{2}}\textbf{\textit{2}}}=-\Gamma_{\textbf{\textit{2}}}\rho_{\textbf{\textit{2}}\textbf{\textit{2}}}+\Gamma_{\textbf{\textit{2}}}\rho_{EE}+J^a_{12}\text{Re}[\rho_{\textbf{\textit{2}}\textbf{\textit{1}}}]\,,\label{tripdmi}\\
&\dot{\rho}_{GG}=\Gamma_{\textbf{\textit{2}}}\rho_{\textbf{\textit{2}}\textbf{\textit{2}}}+\Gamma_{\textbf{\textit{1}}}\rho_{\textbf{\textit{1}}\textbf{\textit{1}}}\,,\\
    &2\text{Re}[\dot{\rho}_{\textbf{\textit{2}}\textbf{\textit{1}}}]=-J^a_{12}(\rho_{\textbf{\textit{2}}\textbf{\textit{2}}}-\rho_{\textbf{\textit{1}}\textbf{\textit{1}}})-(\Gamma_{\textbf{\textit{1}}}+\Gamma_{\textbf{\textit{2}}})\text{Re}[\rho_{\textbf{\textit{2}}\textbf{\textit{1}}}]\,.\label{transition}
\end{align}
For a reciprocal system, i.e., $J^a_{12}=0$, we can solve Eq.~(\ref{transition}) with the solution $\text{Re}[\rho_{\textbf{\textit{2}}\textbf{\textit{1}}}(t)]=\text{Re}[\rho_{\textbf{\textit{2}}\textbf{\textit{1}}}(0)]e^{-(\Gamma_{\textbf{\textit{1}}}+\Gamma_{\textbf{\textit{2}}})t/2}$. Starting from a fully excited state with $\text{Re}[\rho_{\textbf{\textit{2}}\textbf{\textit{1}}}(0)]=0$, we find $\text{Re}[\rho_{\textbf{\textit{2}}\textbf{\textit{1}}}(t)]=0$, which implies the collective state $
\ket{\textbf{\textit{1}}}$ and $
\ket{\textbf{\textit{2}}}$ are decoupled from each other.
Instead, for $J^a_{12}\neq 0$,  the net probability flowing from the symmetric state, i.e., $\ket{\textbf{\textit{2}}}$, to the anti-symmetric state, i.e., $\ket{\textbf{\textit{1}}}$, can be found from Eqs.~(\ref{singdmi}) and~(\ref{tripdmi}) as 
\begin{align}
\Delta P_{\textbf{\textit{1}}\textbf{\textit{2}}}=-2J_{12}^a\text{Re}[\rho_{\textbf{\textit{2}}\textbf{\textit{1}}}]=2J_{\textbf{\textit{1}}\textbf{\textit{2}}}^a\text{Re}[\rho_{\textbf{\textit{2}}\textbf{\textit{1}}}]\,,
\end{align}
which is in agreement with Eq.~(\ref{netproba}). In the following, we show that the coupling between the single excitation states $\ket{\textbf{\textit{1}}}$ and $\ket{\textbf{\textit{2}}}$  results into a spatially asymmetric emission pattern. To this end, we express the emission rate of the $\alpha$th emitter~\eqref{indiviemimod} as 
\begin{align}
\mathcal{R}_{\alpha}(t)=-\sum_{\mu,\nu}\dot{\rho}_{\mu\nu}(t)\bra{\nu}\sigma_{\alpha}^z/2\ket{\mu}\,,
\label{eq26}
\end{align}
 where $\mu,\nu$ represent the collective states $\{G,\textbf{\textit{1}},\textbf{\textit{2}},E\}$. By substituting Eqs.~(\ref{eqone})-(\ref{transition}) into Eq.~\eqref{indiviemimod} we find
\begin{align}
    \mathcal{R}_{1,2}(t)
    =\frac{1}{2}\left[\pm(\dot{\rho}_{\textbf{\textit{1}}\textbf{\textit{2}}}+\dot{\rho}_{\textbf{\textit{2}}\textbf{\textit{1}}})-(\dot{\rho}_{EE}-\dot{\rho}_{GG})\right]\,,\label{emissioinonetwo}
\end{align}
and, correspondingly, Eq.~\eqref{259} becomes
\begin{align}
    \Delta\mathcal{R}_{12}(t)=2\text{Re}[\dot{\rho}_{\textbf{\textit{2}}\textbf{\textit{1}}}(t)]\,.
\end{align}
From Eq.~(\ref{transition}), we see that, when  $J^a_{12}=0$, the real part of $\rho_{\textbf{\textit{2}}\textbf{\textit{1}}}$ remains zero. As a result, $\Delta\mathcal{R}_{12}(t)=0$, implying that the two qubits will emit identically.
\begin{figure*}[htbp]
    \centering
\includegraphics[width=1\linewidth]{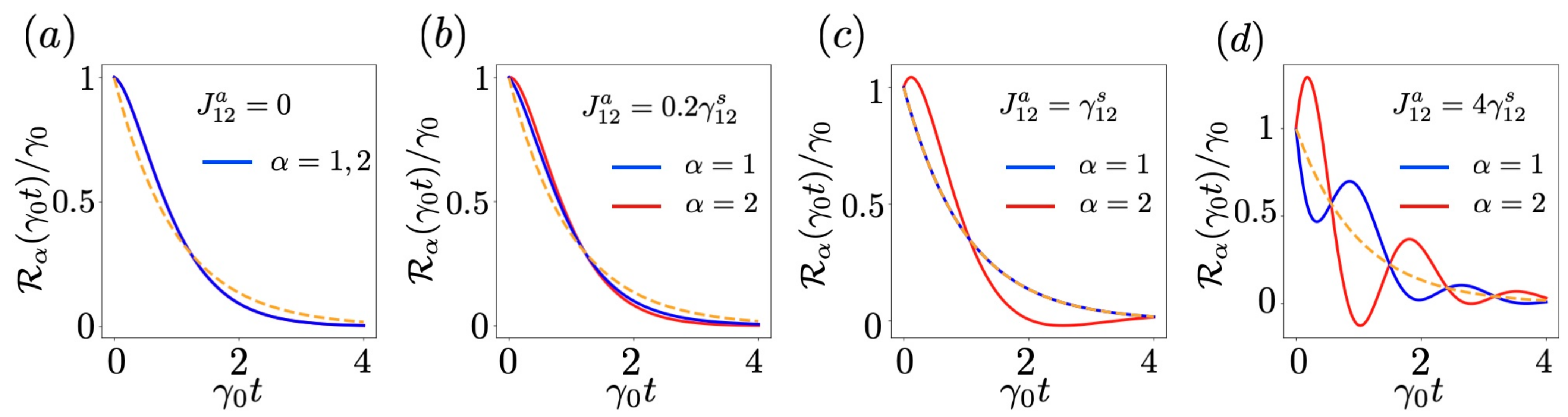}
    \caption{
   Individual emission rates~\eqref{indiviemimod} of two qubits obeying Eq.~(\ref{emissioinonetwo})  as a function of time for increasing strengths of the coherent anti-symmetric interaction $J^{a}_{12}$. The orange dashed line illustrates the relaxation dynamics of an uncorrelated two-qubit ensemble.
  (a) For a reciprocal array, i.e.,  $J^{a}_{12}=0$, the qubit relaxation dynamics are symmetric upon the exchange of the two qubits (blue line). Compared to the decay of an uncorrelated two-qubit ensemble (orange dashed line), the decay rate is initially enhanced and then suppressed at longer times. (b) A weak anti-symmetric interaction \( J_{12}^a = 0.2 \gamma^s_{12} \) introduces a slight asymmetry in the relaxation dynamics of the two qubits; specifically, the emission rate of the right qubit  (\(\alpha = 2\)) at shorter (longer) times is enhanced (suppressed) relative to one of the left qubit (\(\alpha = 1\)). (c) When the inter-qubit interaction becomes unidirectional, corresponding to  $J_{12}^a=\gamma^s_{12}$, the right qubit exhibits a superradiant peak followed by a subradiant tail, while the left qubit decays similarly to an isolated qubit.  (d) For stronger anti-symmetric interactions, e.g., $J_{12}^a=4\gamma^s_{12}$, both qubits display an oscillating decay, albeit out of phase. (a)-(d) In each figure, the time $t$ is expressed in units of the relaxation rate $\gamma_{0}$ of an isolated qubit, and we set $J^{s}_{12}=0$ for simplicity.}
    \label{Fig3}
\end{figure*}

Next, we solve numerically  Eqs.~(\ref{eqone})-(\ref{transition}) for  $J_{12}^{a}=0,0.2 \gamma^s_{12},  \gamma^s_{12}, 4 \gamma^{s}_{12} $  and plot the corresponding emission rates $\mathcal{R}_{1,2}(\gamma_0t)$  in Fig.~\ref{Fig3}(a)-(d). Figures~\ref{Fig3}(a) and (b) show, respectively, that for reciprocal inter-qubit interactions the relaxation dynamics are symmetric upon the exchange of the two qubits, while even a weak non-reciprocity, i.e.,  $J_{12}^a=0.2 \gamma^s_{12}$, can introduce spatial asymmetry in the relaxation dynamics. The most dramatic effect of the anti-symmetric coherent interactions on the asymmetry of the qubit relaxation pattern is observed when the unidirectionality condition  $J_{12}^a=\gamma^s_{12}$, for which $\gamma^{L}_{12}=0$~\eqref{leftcoe}, is satisfied. As displayed by Fig.~\ref{Fig3}(c), under this condition, the relaxation dynamics of the right qubit exhibit a super-radiant peak followed by a sub-radiant tail, while the decay of the left qubit resembles the spontaneous decay of an isolated qubit.
 Finally, for larger values of the anti-symmetric interaction $J_{12}^{a}$, Fig.~\ref{Fig3}(d) shows that the dynamics of both qubits oscillates, albeit out of phase, between super-radiant peaks and sub-radiant valleys around the trajectory of single emitter spontaneous decay, before ultimately relaxing to the ground state.

\section{Solid-state baths} \label{SSbath}
In this Section, we identify the properties of a solid-state reservoir that are essential to the generation of non-reciprocal inter-qubit interactions. 
Following the approach of Ref.~\cite{li2023solid}, we consider an ensemble of identical solid-state spin defects, modelled as two-level systems with the transition $\omega_{qi}$  that interact with the magnetic field $\textbf{B}$ emitted by a nearby stationary solid-state bath. The interaction between the spin defects and the shared bath can be explicitely written as \begin{align}
\mathcal{H}_{SE}=-\tilde{\gamma}\sum_{\alpha=1}^N(B_\alpha^+\sigma^-_\alpha+B_\alpha^-\sigma^+_\alpha+B_\alpha^z\sigma^z_\alpha )\,,\label{bathfield}
\end{align} 
 where $\tilde{\gamma}$ is the gyromagnetic ratio of the 
 solid-state spin defects, and  $B_\alpha^{\pm} = 
 B_\alpha^x \pm iB_\alpha^y$ and $B_\alpha^z$ represent the transverse and longitudinal components, respectively, of the local stray field experienced by the $\alpha$th spin. Since the Zeeman interaction~\eqref{bathfield} is typically much weaker than the characteristic energy scales of the individual systems, the evolution of the density matrix $\rho$ of the qubit ensemble can be generally described by the following Markovian master equation~\cite{li2023solid}:
\begin{align}
    \frac{d\rho}{dt}= -i \left[ \mathcal{H},\;\rho \right] + \mathcal{L}\left[ \rho\right]\,,\label{densitymatrix}
\end{align}
where the Hamiltonian $\mathcal{H}$ reads as 
\begin{align}
\mathcal{H}=  \sum_{\alpha,\beta}\sum_{(\mu,\tilde{\mu})}\sum_{(\nu,\tilde{\nu})}Y^{\mu\nu}_{\alpha\beta} \sigma_\alpha^{\tilde{\mu}}\sigma_\beta^{\tilde{\nu}}\,,\label{master:hamil}
\end{align}
and
\begin{align}
\mathcal{L}\left[ \rho\right]= \sum_{\alpha,\beta}\sum_{(\mu,\tilde{\mu})}\sum_{(\nu,\tilde{\nu})}X_{\alpha\beta}^{\mu\nu}\left( \sigma_\beta^{\tilde{\nu}}\rho\sigma_\alpha^{\tilde{\mu}}-\frac{1}{2}\{ \sigma_\alpha^{\tilde{\mu}}\sigma_\beta^{\tilde{\nu}},\rho\}\right)\label{master:linb}
\end{align}
is the Lindbladian. Here, the sums over $(\mu, \tilde{\mu})$ and $(\nu, \tilde{\nu})$ are taken over the combinations $(\mu, \tilde{\mu}) = (\pm, \mp), (z, z)$. The explicit forms of  $Y^{\mu\nu}_{\alpha\beta}$ and $X^{\mu\nu}_{\alpha\beta}$ are given by~\cite{li2023solid}

\begin{align}
Y^{\mu\nu}_{\alpha\beta}&=\frac{\tilde{\gamma}^2}{2}\int_0^\infty d\tau \; G^>_{B_{\alpha}^\mu B_{\beta}^\nu}(\tau) e^{-i\omega_{qi}^\nu\tau}\nonumber\\
&-\frac{\tilde{\gamma}^2}{2}\int_0^\infty d\tau \; G^<_{B_{\beta}^\nu B_{\alpha}^\mu }(\tau) e^{-i\omega_{qi}^{\mu}\tau}\,,\label{coherentcop}\\
X^{\mu\nu}_{\alpha\beta}&=i\tilde{\gamma}^2\int_0^\infty d\tau \; G^>_{B_{\alpha}^\mu B_{\beta}^\nu}(\tau) e^{-i\omega_{qi}^\nu\tau}\nonumber\\
&+i\tilde{\gamma}^2\int_0^\infty d\tau \; G^<_{B_{\beta}^\nu B_{\alpha}^\mu }(\tau) e^{-i\omega_{qi}^{\mu}\tau}\,,\label{decoherentcop}
\end{align}
where $\omega_{{qi}}^{\pm}=\pm\omega_{{qi}}$ and $\omega_{{qi}}^z=0$.
 Invoking the rotating-wave approximation~\cite{breuer2002theory} and neglecting dephasing effects, which are outside of the scope of this work, Eqs.~\eqref{master:hamil} and~\eqref{master:linb} reduce, respectively, to Eqs.~\eqref{Hmasterfull} and~\eqref{Lmasterfull}. We find that, in the zero-temperature limit, i.e., $T \rightarrow 0$, the coherent and dissipative inter-qubit couplings can be rewritten as

\begin{align}
    2J^s_{\alpha\beta}(\omega_{qi})/\tilde{\gamma}^2
    &=-\text{Re}[\chi^{-+}_{\alpha\beta}(\omega_{qi})]-\text{Re}[\chi^{-+}_{\beta\alpha}(\omega_{qi})]\,, \label{Js}\\
    2J^a_{\alpha\beta}(\omega_{qi})/\tilde{\gamma}^2
    &=-\text{Im}[\chi^{-+}_{\alpha\beta}(\omega_{qi})]+\text{Im}[\chi^{-+}_{\beta\alpha}(\omega_{qi})]\,,\label{Jas}\\
 \gamma^s_{\alpha\beta}(\omega_{qi})/\tilde{\gamma}^2
    &=\,\,\,\,\,\text{Im}[\chi^{-+}_{\alpha\beta}(\omega_{qi})]+\text{Im}[\chi^{-+}_{\beta\alpha}(\omega_{qi})]\,,\label{gammas}\\
   \gamma^a_{\alpha\beta}(\omega_{qi})/\tilde{\gamma}^2
    &=-\text{Re}[\chi^{-+}_{\alpha\beta}(\omega_{qi})]+\text{Re}[\chi^{-+}_{\beta\alpha}(\omega_{qi})]\,, \label{gammaas}
\end{align}
where we assumed that, in the fully excited state, the quantum spins point along the $\hat{z}$ direction. In the linear response regime, the susceptibility of the reservoir is defined as~\cite{kubo1966fluctuation,nakano1993linear,kubo1957statistical,onsager1931reciprocal}  
\begin{align}
    \chi^{-+}_{\alpha \beta}(\omega_{qi})&=i\int_0^{\infty}d\tau\,e^{i\omega_{qi}\tau}\langle[B^{-}_\alpha(\tau),B_\beta^{+}]\rangle\,,\label{SuscepKubo}
\end{align}
where its
 real part, $\text{Re} [\chi^{-+}_{\alpha\beta}(\omega_{qi})]$, and imaginary part, $\text{Im}[\chi^{-+}_{\alpha\beta}(\omega_{qi})]$,  are related by the Kramer-Kronig relations~\cite{toll1956causality}. 
In Sec.~\ref{model}, we have shown that non-reciprocal inter-qubit interactions can be generated via the balancing of  coherent anti-symmetric (symmetric) $J^a_{\alpha\beta}$ ($J^s_{\alpha\beta}$) and dissipative symmetric (anti-symmetric) $\gamma^{s}_{\alpha\beta}$ ($\gamma^{a}_{\alpha\beta}$) interactions. 
From Eqs.~\eqref{Jas} and~\eqref{gammaas}, it is easy to see that the necessary condition for the generation of a non-reciprocal quantum dynamics is a non-reciprocal bath response, i.e., $\chi^{-+}_{\alpha\beta}(\omega_{qi}) \neq \chi^{-+}_{\beta \alpha}(\omega_{qi})$. In light-matter interfaces, this condition translates into the photonic bath lacking time-reversal symmetry~\cite{metelmann2015nonreciprocal}. For a solid-state bath, crystalline symmetries play also a crucial role. Specifically, non-reciprocal linear responses require the simultaneous breaking of both inversion and time-reversal symmetries at the level of the single-particle Hamiltonian. 
Notably, there is a plethora of solid-state systems whose fluctuations can generate electromagnetic noise and inherently exhibit non-reciprocal dynamics.  In the following Section, we explore a concrete example and its potential for realizing non-reciprocal quantum correlations and as a testbed for a novel quantum sensing modality.

\begin{figure*}[htbp]
\centering
\includegraphics[width=1\linewidth]{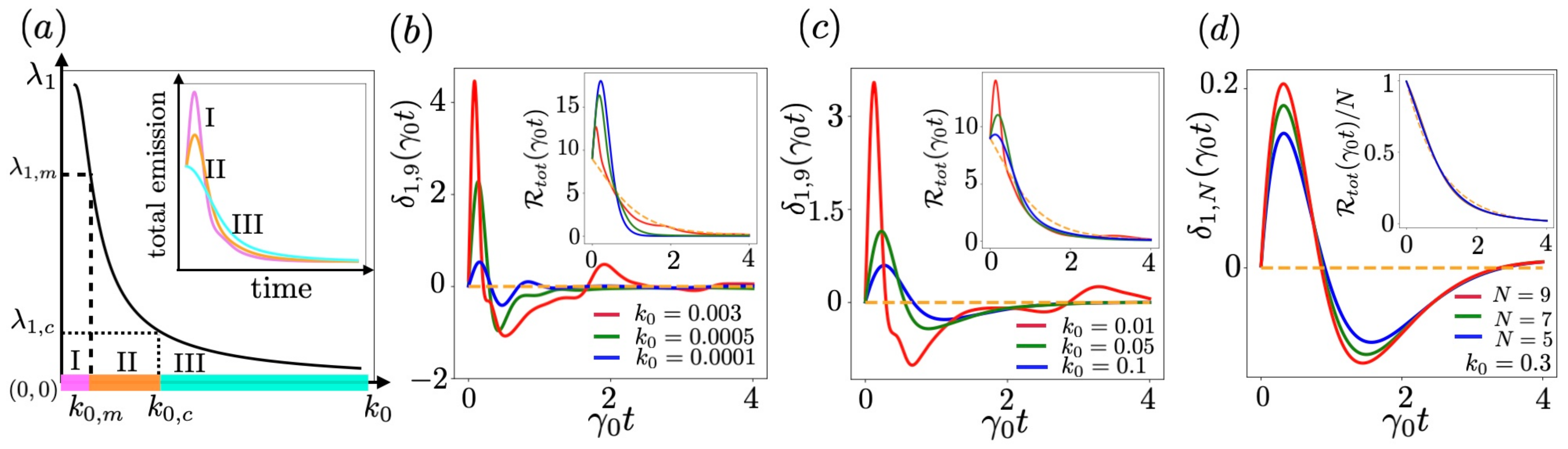}
    \caption{(a) The dependence of the characteristic wavelength $\lambda_1$ on the bath non-reciprocity $k_0$ delineates three regimes in which the degree of non-reciprocity $\delta_{1,N}(\gamma_0 t)$ in the relaxation rate of an ensemble of $N$ qubits initialized in the fully excited state exhibits distinct trends as $k_0$ increases. In Region I, $\delta_{1,N}(\gamma_0 t)$ grows with increasing $k_0$ until it reaches a maximum at $(k_{0,m}, \lambda_{1,m})$, while in Regions II and III, $\delta_{1,N}(\gamma_0 t)$ diminishes as $k_0$ becomes larger.
(b)-(c) The spatial asymmetry in the emission, i.e., $\delta_{1,9}(\gamma_0 t)$, of an ensemble of $N=9$ NV centers initialized in the fully excited state as a function of time for values of $k_0$ corresponding to the regimes I and II defined in (a).
(d) Non-reciprocal emission rate $\delta_{1,N}(\gamma_0 t)$ of an ensemble of $N=5,7,9$ NV centers with $k_0 = 0.3$.
(b)-(d) The dashed orange curves correspond to the relaxation dynamics of $N=9$ isolated NV centers serving as a reference. Insert in (a)-(d): the corresponding total emission rate $\mathcal{R}_{tot}(\gamma_0 t)$ as a function of time.}
    \label{Fig4}
\end{figure*}

\section{Ferromagnetic bath }
\label{concretefeeo}
In this Section, we investigate the asymmetric relaxation patterns that can be observed in a realistic experimental setup and their dependence on both material-specific and experimentally tunable parameters. Several solid-state systems, ranging from non-centrosymmetric crystals to specially designed magnetic heterostructures, display non-reciprocal responses~\cite{nagaosa2024nonreciprocal,cheong2018broken,tokura2018nonreciprocal,sato2019nonreciprocal,sato2016magnon,dhital2017exploring}.
Here, we take as a case study a generic non-centrosymmetric ferromagnetic system with chiral Dzyaloshinskii–Moriya (DM) interactions and explore how its spin density fluctuations affect the dynamics of a nearby qubit ensemble.

Specifically, we consider a NV-center array with lattice constant $a_{q}$ arranged along the $x$ direction in a plane at a distance $d$ above the magnetic bath. While a NV center forms a spin-triplet system in the ground state, the degeneracy between $m_s=\pm1$ can be lifted by  applying an external magnetic field $B_{0}$ along its principal axis, here taken to be the $\mathbf{\hat{z}}$ axis~\cite{jelezko2006single,childress2013diamond}.  Each qubit can then be treated as a two-level system with frequency $\omega_{qi}=\Delta_{0}-\tilde{\gamma} B_{0}$, where $\Delta_0=2.87\ \text{GHz}$ is the NV-center zero-field splitting. This approximation is valid as long as the thermal energy of the system is significantly smaller than the energy of the higher NV-center transition, i.e., $k_{B}T \ll \Delta_{0}+\tilde{\gamma} B_{0}$, which is consistent with the quantum regime (i.e., $T\rightarrow 0$) we are focusing on. At the position $\mathbf{r}_{\alpha}$ of the $\alpha$th NV center, the local spin density $\mathbf{s} (\mathbf{r})$ of the magnetic film with thickness $t_F$ generates a magnetic field with components given by
\begin{align}
    B^{\mu}_\alpha=\gamma\,t_F\int\,d^2\mathbf{r}\;\mathcal{D}^{\mu\nu}(\mathbf{r}_\alpha-\mathbf{r})s^{\nu}(\mathbf{r})\,,\label{fieldtospin}
\end{align}
where $\gamma$ is the gyromagnetic ratio of the film and $\mathcal{D}^{\mu\nu}(\mathbf{r}_\alpha-\mathbf{r})$ is the tensorial Green’s function of the magnetic dipolar field, which can be obtained as solutions of the Maxwell's equations   in the magnetostatic limit~\cite{guslienko2011magnetostatic}. Here, the repeated indices $\mu,\nu\in\{+,-,z\}$ are implicitly summed over.  By plugging Eq.~\eqref{fieldtospin} into Eq.~\eqref{SuscepKubo}, one finds
\begin{align}
    \chi^{-+}_{\alpha\beta}(\omega_{qi})=\gamma^2t_F\int\,d^2\mathbf{k} \; k^2e^{-2kd}e^{i\mathbf{k}\cdot\mathbf{r}_{\alpha\beta}}\chi_{s^-s^+}(\omega_{qi},\mathbf{k})\,,\label{suctospinsuc}
\end{align}
where
\begin{align}
     \chi_{s^\mu\,s^\nu}(\omega_{qi},\mathbf{k})=\int_0^\infty\,d\tau\,e^{i\omega_{qi}\tau}(i\langle [s^\mu_{\mathbf{k}}(\tau),s^\nu_{-\mathbf{k}}]\rangle)\,\label{quantumsus}
 \end{align}
 is the film spin susceptibility.
The latter, in turn, depends on the spin-wave dispersion, which we take as
\begin{align}
 \omega_F(\mathbf{k})=J k^2 - 2D k_x +\Delta\,,
 \label{dispersion}
 \end{align}
 where $\mathbf{k}=(k_x,k_y)$, with $ |\mathbf{k}|=k$, is the (dimensionless) spin-wave momentum,  $J$ and $D$ (in units of frequency) parameterize, respectively, the symmetric and anti-symmetric exchange interactions, and   $\Delta$ is a local term  dependent on the field $B_{0}$, which enforces a collinear out-of-plane spin alignment and accounts for the magnetocrystalline anisotropies intrinsic to the ferromagnetic film. The spin-wave dispersion~\eqref{dispersion} is clearly non-reciprocal, i.e., $\omega_F(\mathbf{k})\neq\omega_F(-\mathbf{k})$. Here, the global inversion symmetry of the system is broken by the non-centrosymmetric crystalline structure. The lack of inversion symmetry, in turn, allows for finite DM interactions $\propto D$, which break the effective time-reversal symmetry of the magnon Hamiltonian~\cite{dzyaloshinsky1958thermodynamic,moriya1960new,moriya1960anisotropic}.

 It is important to note that Eq.~\eqref{dispersion} does not account for the microscopic details of a specific magnetic system. Nonetheless, it serves as a minimal model that allows us to analyze our results as a function of the parameter $D/J$, which provides a measure of non-reciprocity that can be easily related to more complex, material-specific models. For concreteness, we set the field to $B_{0} = 400$ G  and we take the other parameters akin to the ones of a centrosymmetric Yttrium Iron Garnet (YIG) film of thickness $t_F=20$~nm, spin density $s=1.2\times10^{-10}\;\text{G}^2\cdot\text{cm}\cdot\text{s}$, zero-field gap $\Delta (B_0=0) = 0.55$ GHz and $J=\rho_{s}/a^2$, where $\rho_{s} = 7.7\times10^{-6}\;\text{Hz}\cdot\text{m}^2$ is the spin stiffness and $a=1.2$~nm is the lattice constant. 

Using Eq.~\eqref{dispersion}, we can rewrite Eq.~(\ref{quantumsus}) as
  \begin{align}
   \chi_{s^-s^+}(\omega_{qi},\mathbf{k})=P.V.\frac{s/2J}{\Omega(\mathbf{k})}+i\frac{\pi\,s}{2J}\delta(\Omega(\mathbf{k}))\,,\label{spinsussamllgilbert}
\end{align}
where $P.V.$ denotes the principle value and
\begin{align}
    \Omega(\mathbf{k})=[\omega_F(\mathbf{k})-\omega_{qi}]/J=\left(\mathbf{k}-k_0\hat{\mathbf{x}}\right)^2+k_1^2\,,\label{spectrum}
\end{align}
with  $k_0=D/J$  and $k_1=\sqrt{k_0^2+(\omega_{qi}-\Delta)/J}$.  
By plugging Eqs.~(\ref{suctospinsuc}), (\ref{spinsussamllgilbert}) and (\ref{spectrum}) into Eqs.~(\ref{Js})-(\ref{gammaas}), we can systematically derive the coherent $J_{\alpha \beta}$ (\ref{Hmasterfull}) and 
dissipative $\gamma_{\alpha\beta}$ (\ref{Lmasterfull}) inter-qubit interactions.  Although the resulting expressions involve complicated integrals that cannot be solved analytically, one can gain physical insights by taking the limit $e^{d / \lambda_1} \rightarrow 1$,  which allows us to write 
\begin{align}
    \frac{\gamma_{\alpha\beta}}{\gamma_0}&=e^{i\frac{x_{\alpha\beta}}{\lambda_0}}\left[J_0\left(\frac{x_{\alpha\beta}}{\lambda_1}\right)+i\frac{2\lambda_0\lambda_1}{\lambda_0^2+\lambda_1^2}J_1\left(\frac{x_{\alpha\beta}}{\lambda_1}\right)\right],\label{gammamag}\\
    \frac{J_{\alpha\beta}}{\gamma_0}&=e^{i\frac{x_{\alpha\beta}}{\lambda_0}}\left[Y_0\left(\frac{x_{\alpha\beta}}{\lambda_1}\right)+i\frac{2\lambda_0\lambda_1}{\lambda_0^2+\lambda_1^2}Y_1\left(\frac{x_{\alpha\beta}}{\lambda_1}\right)\right]\label{Jmag},
\end{align}
where $\lambda_{0,1} = a/k_{0,1}$, $\gamma_0=\frac{2\pi^2\,h^2 (\tilde{\gamma}\gamma)^2t_F\,s\lambda_0^2}{\rho_{s} \lambda_1^2(\lambda_0^2+\lambda_1^2)}$ and $J_{0,1}(\cdots)$ and $Y_{0,1}(\cdots)$ are the Bessel functions of the first and second kind, respectively.  It is easy to verify that, for a reciprocal reservoir, i.e., $k_{0} \rightarrow0$, Eqs.~(\ref{gammamag}) and (\ref{Jmag}) reduce to purely real inter-qubit interactions discussed  in Ref.~\cite{li2023solid} for which $J^{a}_{\alpha\beta},\gamma^{a}_{\alpha\beta}=0$. In this limit, the  relaxation dynamics of an array of $N$ qubits are spatially symmetric, i.e., $\mathcal{R}_{\alpha}(\gamma_0 t) = \mathcal{R}_{\small \alpha'}(\gamma_0 t)$ with $\alpha'=(N+1)-\alpha$, and the  collective emission rate $\mathcal{R}_{tot}(\gamma_0 t)=\sum_{\alpha=1}^N\mathcal{R}_{\alpha}(\gamma_0t)/\gamma_0$   displays a super-radiant burst followed by sub-radiant tail  when the characteristic wavelength $\lambda_1$ is larger than a critical value $\lambda_{1,c}$ ~\cite{li2023solid}. 

On the other hand, a finite degree of non-reciprocity in the bath, i.e., $k_{0} \neq 0$, introduces a spatial asymmetry in the array emission pattern, which can be quantified in terms of the difference in the normalized emission rates of the left- and rightmost qubits, i.e., $\delta_{1,N}(\gamma_0 t) = [\mathcal{R}_{1}(\gamma_0 
t) - \mathcal{R}_{N}(\gamma_0 t)] / \gamma_0$.  One might 
intuitively expect the parameter $\delta_{1,N}(\gamma_0 t)$ 
to increase with increasing bath non-reciprocity, i.e., as $k_{0}$ becomes larger. However, as the bath non-reciprocity increases for a qubit array with a fixed lattice constant, the characteristic wavelength $\lambda_1$ controlling the spatial range of inter-qubit interactions decreses, yielding the trend conceptually summarized in Fig.~\ref{Fig4}(a). In region I, the spatial emission asymmetry $\delta_{1,N}(\gamma_0 t)$ in the array increases as the bath non-reciprocity grows, reaching its maximum at $k_0 = k_{0,m}$. 
However, as the bath non-reciprocity increases further, both the height of the super-radiant peak in the collective emission (see inset) and the asymmetry between individual relaxation rates diminish (Region II). This reduction arises from the fact that the characteristic wavelength \(\lambda_1\), which governs the strength of inter-qubit interactions, decreases with increasing \(k_0\). Eventually, at \(k_{0} = k_{0,c}\), the characteristic wavelength reaches the critical value \(\lambda_{1} = \lambda_{1,c}\), at which point the super-radiant peak in the collective emission vanishes. Figures~\ref{Fig4}(b) and (c) show explicitly these trends for an array of $N=9$ NV centers with lattice spacing  $a_q = 20$ nm  initialized in the fully excited state. The asymmetry $\delta_{1,N}(\gamma_0 t)$  between the emission rates of the left- and rightmost qubits reaches its maximum value at approximately $k_{0,m}=0.003$ (red curve in Fig.~\ref{Fig4}(b), and then decreases as the bath non-reciprocity increases further. The (normalized) collective emission rate $\mathcal{R}_{tot}(\gamma_0t)$  displayed in the insets instead decreases monotonically as $k_{0}$ increases, and the signatures of super- and sub-radiant behaviors in the emission rate vanish when $k_{0,c} \sim0.2$.
On the other hand, for larger values of \(k_0\), while the super- and sub-radiant signatures disappear, the asymmetry in the relaxation dynamics still persists, albeit weakened compared to Regions I and II. Furthermore, the latter still exhibits a cooperative nature, i.e., \(\delta_{1,N}(\gamma_0 t)\) increases with the number \(N\) of qubits, as shown in Fig.~\ref{Fig4}(d). 

Our results indicate that the symmetry breaking should be detectable within a broad range of parameters provided the NV-center ensemble is sufficiently dense. While exploiting bath non-reciprocity for building non-reciprocal quantum circuits would necessitate individual qubit addressability, we propose that the symmetry of the bath can be probed by comparing the collective relaxation rates of qubit sub-ensembles located at the left and right edges of a dense ensemble. This insight is further supported by experimental evidence showing that the relaxation dynamics of an NV-center ensemble, interacting dissipatively with a bosonic reservoir, can exhibit clear signatures of nonlocal correlations, even in the presence of strong disorder and broadening, when the ensemble is highly dense~\cite{angerer2018superradiant}.

\section{Discussion and conclusions\label{conclusion}}
In this work, we investigate the relaxation dynamics of a qubit ensemble interacting with a bath that generates both symmetric and anti-symmetric coherent and dissipative qubit-qubit couplings. We formally show that the interplay between symmetric (anti-symmetric) coherent and anti-symmetric (symmetric) dissipative interactions gives rise to non-reciprocal emission dynamics, and we further link this phenomenon to the non-reciprocal dynamical evolution of the many-body quantum states. 
Applying our theoretical framework to solid-state reservoirs, we identify the breaking of inversion and time-reversal symmetry as crucial for the emergence of quantum non-reciprocity, and we connect symmetric (anti-symmetric) inter-qubit couplings to symmetrized (anti-symmetrized) bath response functions. While non-reciprocal quantum many-body dynamics can be achieved in quantum optics systems through non-linearities or external drives, our results show that these dynamics can also be realized using stationary solid-state baths that intrinsically display non-reciprocal dynamics.

Turning the tables, we propose that the asymmetry in the relaxation dynamics of an ensemble of quantum sensors can serve as a probe for symmetry breaking in the bath. While spin-defect-based relaxometry has already enabled the probing of a range of solid-state phenomena with unprecedented resolution~\cite{casola2018probing}, solid-state spin qubits have, until now, been operated essentially as single-qubit sensing devices. Quantum sensing schemes that rely on signals from multiple NV centers have also only been implemented at temperatures high enough to render any potential quantum correlations among the sensors irrelevant~\cite{rovny2022nanoscale}. In this context, our work advances the development of quantum sensing approaches that capitalize on the quantum correlations among multiple sensors — an area that remains largely unexplored. To assess the experimental significance of our proposal, we explore the relaxation dynamics of an ensemble of NV centers interacting dissipatively with a non-centrosymmetric magnetic bath and find that its symmetry breaking can be identified by probing the relaxation dynamics of the ensemble over a broad range of experimental parameters.

\section{ACKNOWLEDGEMENTS}

Research on the interplay between non-reciprocal emission and many-body physics was supported by DOE under Award No. DE-SC0024090. Research on its application to quantum sensing was supported by ONR under Award No. N000142412427.

\bibliography{library}

\end{document}